\documentclass[3p,times,twocolumn]{elsarticle}

\usepackage{ecrc}

\volume{00}

\firstpage{1}

\journalname{Nuclear Physics B Proceedings Supplement}

\runauth{O. Brandt for the D0 Collaboration}

\jid{nuphbp}

\jnltitlelogo{Nuclear Physics B Proceedings Supplement}

\usepackage{amssymb}
\usepackage{lineno}

\usepackage[figuresright]{rotating}
\usepackage{graphicx}  % needed for figures
\usepackage{dcolumn}   % needed for some tables
\usepackage{bm}        % for math
\usepackage{amssymb}   % for math
\usepackage[T1]{fontenc}
\usepackage[latin1]{inputenc}
\usepackage{longtable}

\usepackage{url}
\usepackage{xspace}
\usepackage{multirow}
\usepackage{units}
\usepackage[normalem]{ulem}
\usepackage{color}

\newcommand{\ttbar}{\ensuremath{t\bar{t}}\xspace}

\newcommand{\etmiss}{\ensuremath{E \kern-0.6em\slash_{\rm T}}\xspace}
\newcommand{\etmissx}{\ensuremath{E \kern-0.6em\slash_{\rm x}}\xspace}
\newcommand{\etmissy}{\ensuremath{E \kern-0.6em\slash_{\rm y}}\xspace}
\newcommand{\alpgen}{{\sc alpgen}\xspace}
\newcommand{\pythia}{{\sc pythia}\xspace}
\newcommand{\herwig}{{\sc herwig}\xspace}
\newcommand{\vecbos}{{\sc vecbos}\xspace}

\newcommand{\mt}{\ensuremath{m_t}\xspace}
\newcommand{\mw}{\ensuremath{M_W}\xspace}
\newcommand{\kjes}{\ensuremath{k_{\rm JES}}\xspace}
\newcommand{\pevt}{\ensuremath{P_{\rm evt}}\xspace}
\newcommand{\psig}{\ensuremath{P_{\rm sig}}\xspace}
\newcommand{\pbkg}{\ensuremath{P_{\rm bkg}}\xspace}

\newcommand{\ejets}{\ensuremath{e+{\rm jets}}\xspace}
\newcommand{\mujets}{\ensuremath{\mu+{\rm jets}}\xspace}
\newcommand{\ljets}{\ensuremath{\ell+{\rm jets}}\xspace}

\newcommand{\etal}{\textit{et al.}\xspace}

\newcommand{\GeV}{\ensuremath{\textnormal{GeV}}\xspace}

\newcommand{\dif}{\ensuremath{{\rm d}}}

\newcommand{\met}{\ensuremath{{}/\!\!\!{p}_{\rm T}}\xspace}

\newcommand{\fb}{\ensuremath{{\rm fb}^{-1}}\xspace}
\newcommand{\mtop}{\ensuremath{m_t}\xspace}

\newcommand{\pt}{\ensuremath{p_{\rm T}}\xspace}

\newcommand{\x}{\ensuremath{\vec x}\xspace}
\newcommand{\y}{\ensuremath{\vec y}\xspace}

\begin{document}

\begin{frontmatter}



\dochead{}

\title{Measurements of the top quark mass with the D0 detector}


\author{Oleg Brandt (for the D0 Collaboration)}

\address{
Kirchhoff Institut f\"ur Physik, University of Heidelberg\\
\vspace{1cm}
\mbox{\normalsize{\rm FERMILAB-CONF-14-390-E}}
}

\begin{abstract}
The mass of the top quark is a fundamental parameter of the standard model (SM) and has to be determined experimentally. In this talk, I present the most recent measurements of the top quark mass in $p\bar p$ collisions at $\sqrt s=1.96$~TeV recorded by the D0 experiment at the Fermilab Tevatron Collider. The measurements are performed in final states containing two leptons, using 5.4~\fb of integrated luminosity, and one lepton, using 9.7~\fb of integrated luminosity. The latter constitutes the most precise single measurement of the mass of the top quark, corresponding to a relative precision of 0.43\%. I conclude with a combination of our results with the results by the CDF collaboration, attaining a relative precision of 0.37\%.
\end{abstract}


\end{frontmatter}

%\linenumbers


\section{Introduction}
Since its discovery~\cite{bib:discoverydzero,bib:discoverycdf}, the determination of the properties of the top quark has been one of the main goals of the Fermilab Tevatron Collider, recently joined by the CERN Large Hadron Collider. The measurement of the top quark mass \mt, a fundamental parameter of the standard model (SM), has received particular attention.
%attention~\cite{bib:combitev,bib:combitevprd,bib:combilhc}, 
Indeed, \mt, the mass of the $W$ boson \mw, and the mass of the Higgs boson are related through radiative corrections that provide an internal consistency check of the SM~\cite{bib:lepewwg}. Furthermore, \mt dominantly affects the stability of the SM Higgs potential, which has related cosmological implications~\cite{bib:vstab1,bib:vstab2,bib:vstab3}.
Currently, with $\mt=173.34\pm0.76~\GeV$, a world-average combined precision of about 0.5\% has been achieved~\cite{bib:combitevprd,bib:combitev,bib:combiworld}.

Measurements of properties of the top quark other than \mtop\ at D0 are reviewed in Ref.~\cite{bib:proptalk}. The full listing of top quark measurements at D0 can be found in~Refs.~\cite{bib:topresd0}.

At the Tevatron, top quarks are mostly pro~duced in pairs via the strong interaction. %, in about 85\% of the cases via $q\bar q'$ annihilation and in about 15\% via gluon-gluon fusion. 
By the end of Tevatron operation, about 10fb$^{-1}$ of integrated luminosity were recorded by D0, which corresponds to about 80k produced $\ttbar$ pairs. In the framework of the SM, the top quark decays to a $W$~boson and a $b$~quark nearly 100\% of the time, resulting in a $W^+W^-b\bar b$ final state from top quark pair production. 
%One of the challenges in measuring the top quark mass is the assignment of reconstruced leptons, jets, and missing transverse energy $\etmiss$ to partons, which, in absence of jet charge and flavour identification, can lead to several possible combinations. 
Thus, $\ttbar$ events are classified according to the $W$ boson decay channels as ``dileptonic'', ``all--jets'', or ``lepton+jets''.

\section{Measurement of the top quark mass in dilepton final states}
The  most precise determination of \mtop\ in dilepton final states at the Tevatron is performed by D0 using 5.4~\fb\ of data~\cite{bib:mtopll_d0}. It is a combination of two measurements, using the matrix element (ME) technique~\cite{bib:mtopllme_d0}, which will be described in Sec.~\ref{sec:ljets} in the context of the \ljets\ channel, and the neutrino weighting technique~\cite{bib:mtopll_d0}. Leaving \mtop\ as a free parameter, dilepton final states are kinematically underconstrained by two degrees of freedom. To account for this in the analysis using ME, a prior is assumed for the transverse momentum distribution of the \ttbar\ system, and the neutrino momenta are integrated over. In the neutrino weighting analysis, distributions in rapidities of the neutrino and the antineutrino are postulated, and a weight is calculated, which depends on the consistency of the reconstructed $\vec\pt^{\,\nu\bar\nu}\equiv\vec\pt^{\,\nu}+\vec\pt^{\,\bar\nu}$ with the measured missing transverse momentum $\met$ vector, versus \mtop. D0 uses the first and second moment of this weight distribution to define templates and extract \mtop. To reduce the systematic uncertainty, the {\em in situ} JES calibration in \ljets\ final states derived in Ref.~\cite{bib:mtoplj_d0} is applied, accounting for differences in jet multiplicity, luminosity, and detector ageing. A combination of both analyses in the dilepton final states at D0 yields $\mtop=173.9~ \pm 1.9~({\rm stat}) \pm 1.6~({\rm syst})~\GeV$.

\section{Measurement of the top quark mass in lepton+jets final states}
\label{sec:ljets}
The most precise measurement of \mt at D0 is performed in \ljets final state with a matrix element (ME) technique, which determines the probability of observing each event under both the $\ttbar$ signal and background hypotheses described by the respective MEs~\cite{bib:run1nature}.
The overall jet energy scale (JES) is calibrated {\it in situ} by constraining the reconstructed invariant mass of the hadronically decaying $W$ boson to $\mw=80.4$~GeV~\cite{bib:wmass}. The measurement is performed using the full set of $p\bar p$ collision data at $\sqrt s=1.96~$TeV recorded by the D0 detector in the Run II of the Fermilab Tevatron Collider, corresponding to an integrated luminosity of $9.7~\fb$. %This is an update of a previous D0 measurement that used 3.6~\fb of integrated luminosity and measured $\mt=174.94\pm1.14\thinspace({\rm stat+JES})\pm0.96\thinspace({\rm syst})~\GeV$~\cite{bib:mtprd}. 
In the present measurement, we not only use a larger data sample to improve the statistical precision, but also refine the estimation of systematic uncertainties through an updated detector calibration, in particular improvements to the $b$-quark JES corrections~\cite{bib:jes}, and using recent improvements in  modeling the \ttbar signal. The analysis was performed blinded in \mt.

This analysis requires the presence of one isolated electron or muon with  transverse momentum $\pt>20~\GeV$ and $|\eta|<1.1$ or $|\eta|<2$, respectively. In addition, exactly four jets with $\pt>20~\GeV$ within $|\eta|<2.5$, and $\pt>40~\GeV$ for the jet of highest \pt, are required. Jet energies are corrected to the particle level using calibrations derived from exclusive $\gamma+$jet, $Z+$jet, and dijet events~\cite{bib:jes}. These calibrations account for differences in detector response to jets originating from a gluon, a $b$~quark, and $u,d,s,$ or $c$~quarks. 
Furthermore, each event must have an imbalance in transverse momentum of $\met>20~\GeV$ expected from the undetected neutrino.
To further reduce background, at least one jet per event is required to be tagged as originating from a $b$ quark ($b$-tagged).

The extraction of \mt is based on the kinematic information in the event and performed with a likelihood technique using per-event probability densities (PD) defined by the MEs of the processes contributing to the observed events. %, giving the ME technique its name. 
Assuming only two non-interfering contributing processes, \ttbar and $W+{\rm jets}$ production, the per-event PD is:
\begin{eqnarray}
\pevt &=& A(\x)[ f\psig(\x; \mt,\kjes)\nonumber\\
      &+& (1-f)\pbkg(\x;\kjes) ]\,,\label{eq:pevt}
\end{eqnarray}
where the observed signal fraction $f$, \mt, and the overall multiplicative factor adjusting the energies of jets after the JES calibration $\kjes$, are parameters to be determined from data. Here, $\x$ represents the measured jet and lepton four-momenta, and $A(\x)$ accounts for acceptance and efficiencies. 
The function \psig describes the PD for \ttbar production. Similarly, \pbkg describes the PD for $W+{\rm jets}$ production, which contributes 14\% of the data in the \ejets and 20\% in the \mujets channels.

% P_sig definition0
In general, the set $\x$ of measured quantities will not be identical to the set of corresponding partonic variables $\y$ because of finite detector resolution and parton hadronization. Their relationship is described by the transfer function $W(\x,\y,\kjes)$, where we assume that the jet and lepton angles are known perfectly. The densities \psig and \pbkg are calculated through a convolution of the differential partonic cross section, $\dif\sigma(\y)$, with $W(\x,\y,\kjes)$ and the PDs for the initial-state partons, $f(q_i)$, where the $q_i$ are the momenta of the colliding partons, by integrating over all possible parton states leading to $\x$:
\begin{eqnarray}
\psig &=& \frac1{\sigma_{\rm obs}^{\ttbar}(\mt,\kjes)}\int\sum~~\dif\sigma(\y,\mt)\dif \vec q_1\dif\vec q_2 \nonumber\\
&\times& f(\vec q_1)f(\vec q_2) W(\x,\y;\kjes)\,.\label{eq:psig}
\end{eqnarray}
The sum in the integrand extends over all possible flavor combinations of the initial state partons. The longitudinal momentum parton density functions (PDFs), $f(q_{i,{\rm z}})$, are taken from the CTEQ6L1 set~\cite{bib:cteq}, while the dependencies $f(q_{i,{\rm x}})$, $f(q_{i,{\rm y}})$ on transverse momenta are taken from PDs obtained from the \pythia simulation~\cite{bib:pythia}. The factor ${\sigma_{\rm obs}^{\ttbar}(\mt,\kjes)}$, defined as the expected total $\ttbar$ cross section,
%for a given $(\mt,\kjes)$, 
ensures that $A(\x)\psig$ is normalized to unity. The differential cross section, $\dif\sigma(\y,\mt)$, is calculated using the leading order (LO) ME for the process $q\bar q\to\ttbar$. The $M_W = 80.4$~GeV constraint for the {\it in-situ} JES calibration is imposed by integrating over $W$ boson masses from a Breit-Wigner prior.

The density \psig is calculated by numerical Monte Carlo (MC) integration. For this, we utilize the Sobol low discrepancy sequence~\cite{bib:sobol} instead of pseudo-random numbers. This provides a reduction of about one order of magnitude in calculation time. Furthermore, we approximate the exact results of Eq.~(\ref{eq:psig}) for a grid of points in $(\mt,\kjes)$ space by calculating the ME only once for each $\mt$ and multiplying the results with the transfer function $W(\x,\y;\kjes)$ to obtain \psig for any \kjes. This results in another order of magnitude reduction in computation time. Both improvements proved essential to reduce the statistical uncertainty in evaluating most of the systematic uncertainties discussed below.

\begin{figure}
\begin{centering}
\includegraphics[width=0.49\columnwidth]{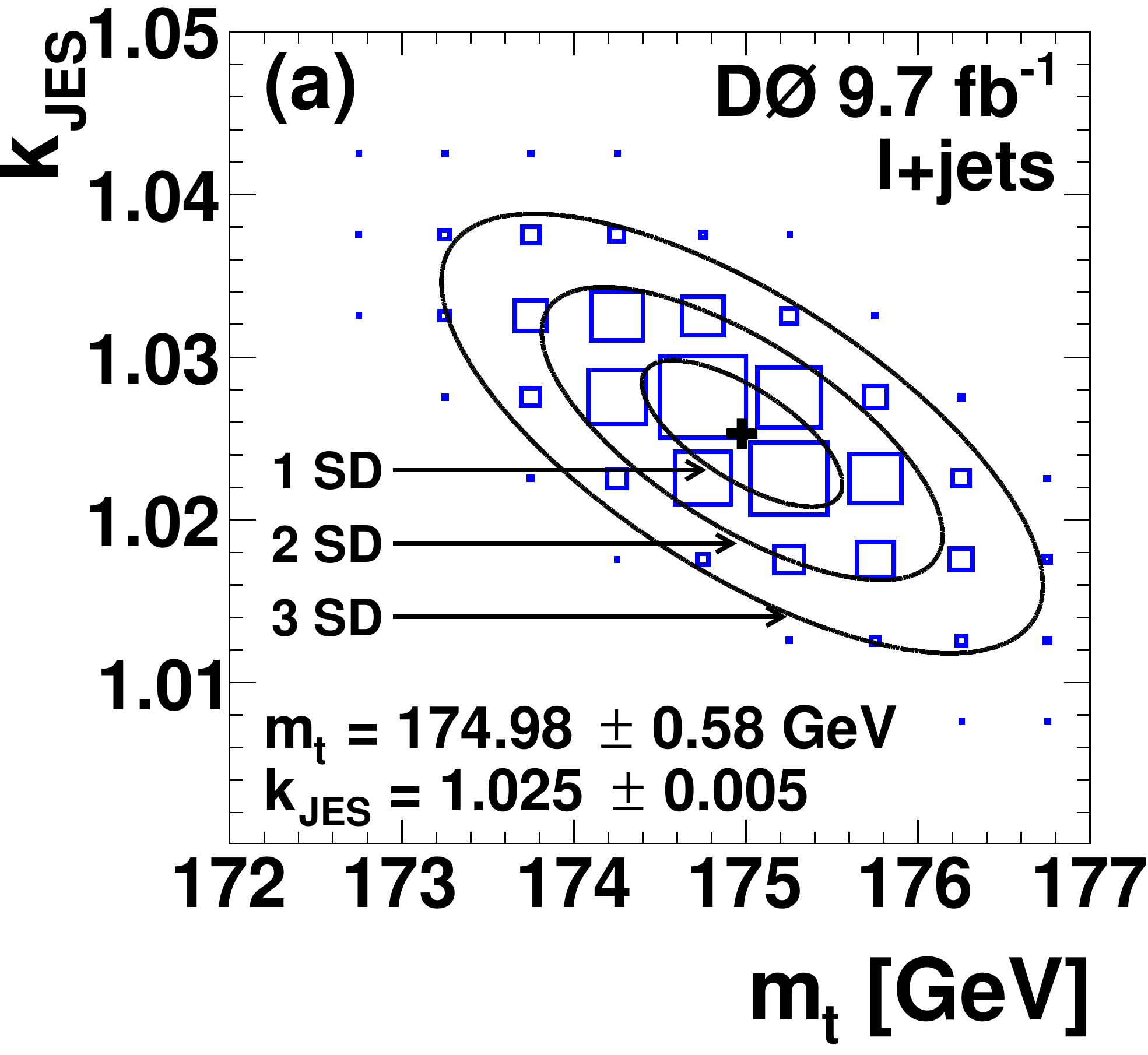}
\includegraphics[width=0.49\columnwidth]{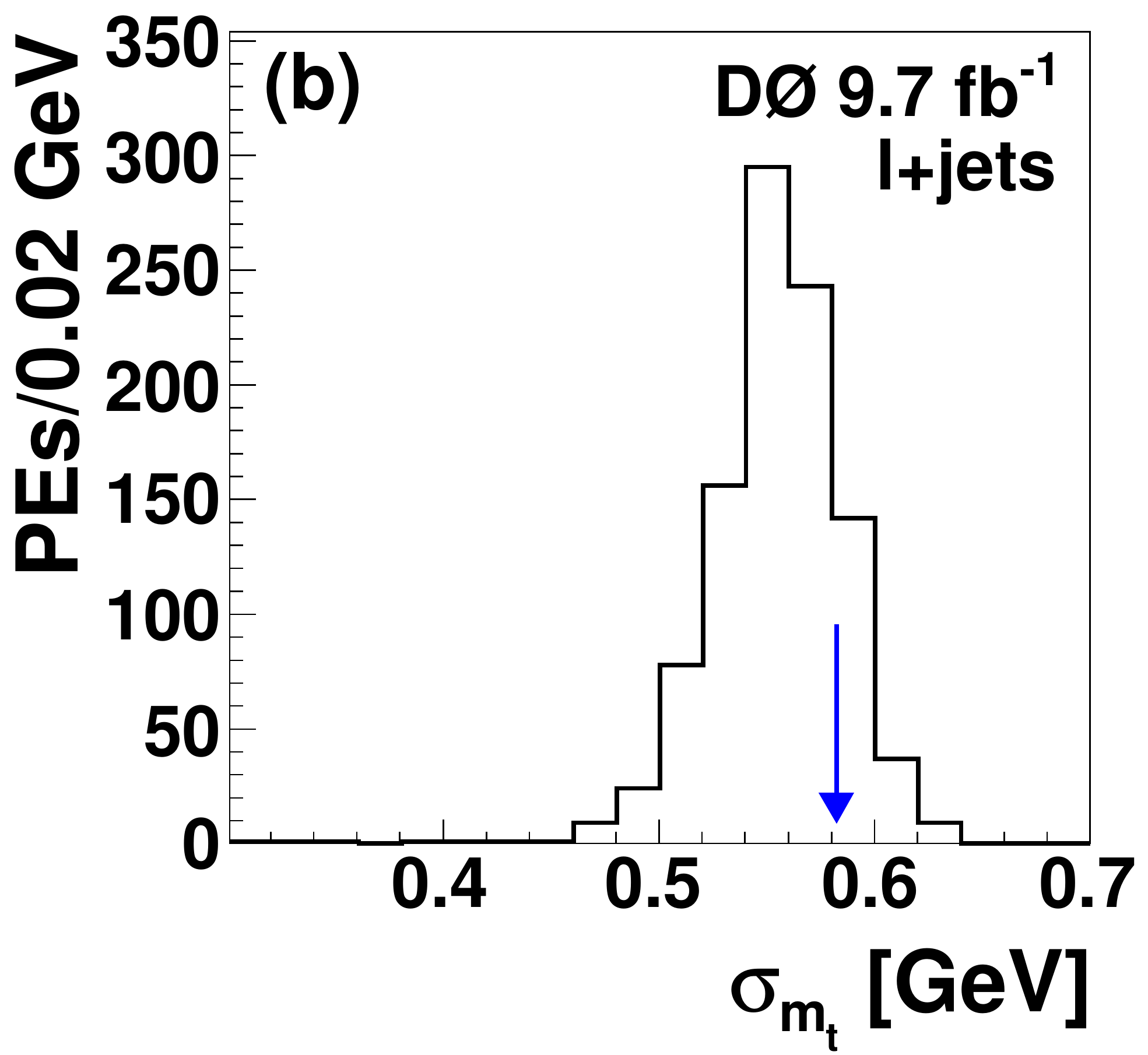}
\par\end{centering}
\caption{
\label{fig:like2d}
(color online) 
(a) Two-dimensional likelihood ${\cal L}(\mt,\kjes)/{\cal L}_{\rm max}$ for data. Fitted contours of equal probability are overlaid as solid lines. The maximum is marked with a cross. Note that the bin boundaries do not necessarily correspond to the grid points on which $\cal L$ is calculated.
(b) Expected uncertainty distributions for \mt with the measured uncertainty indicated by the arrow.
}
\end{figure}

\begin{figure}
\begin{centering}
\includegraphics[width=0.49\columnwidth]{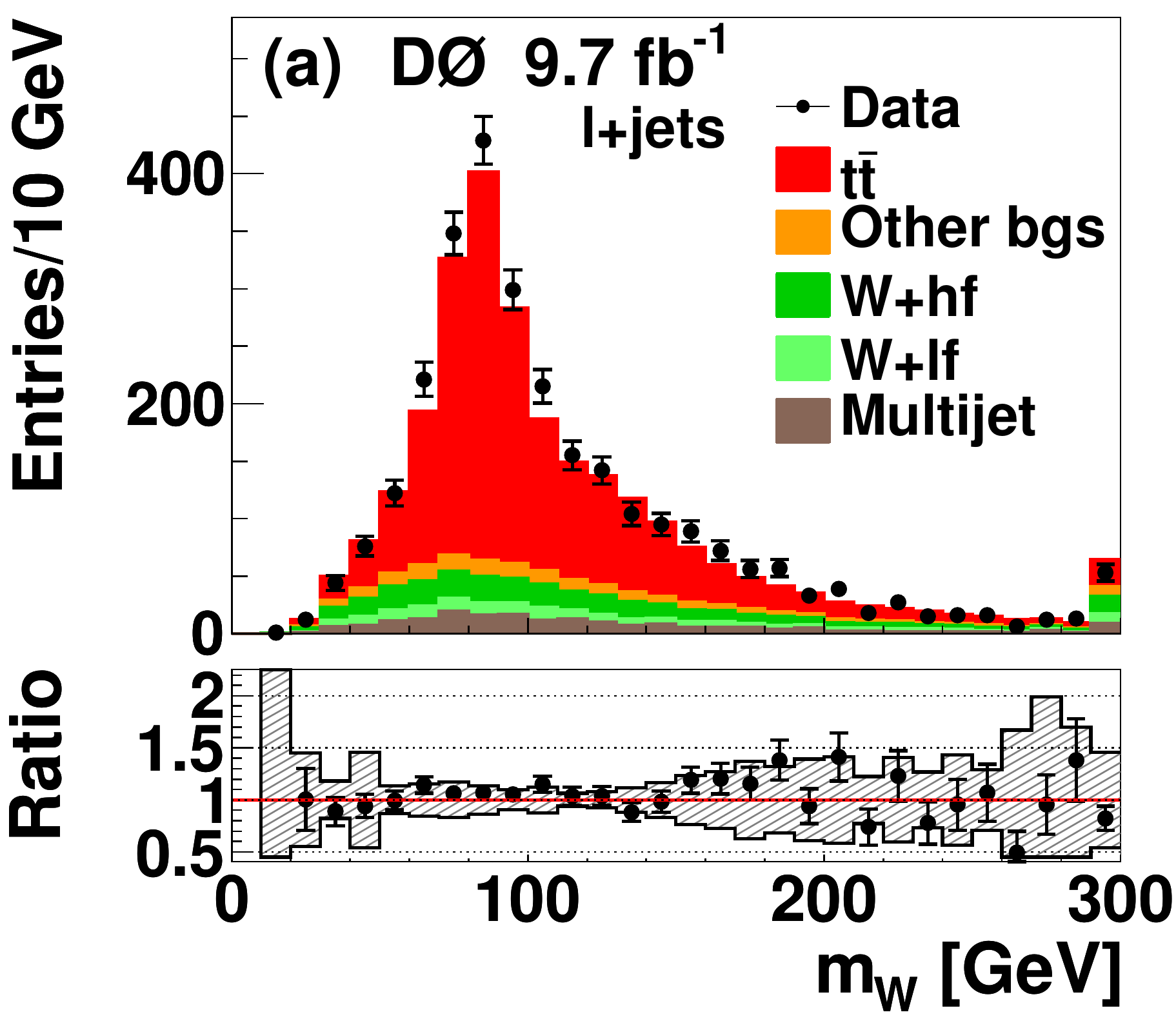}
\includegraphics[width=0.49\columnwidth]{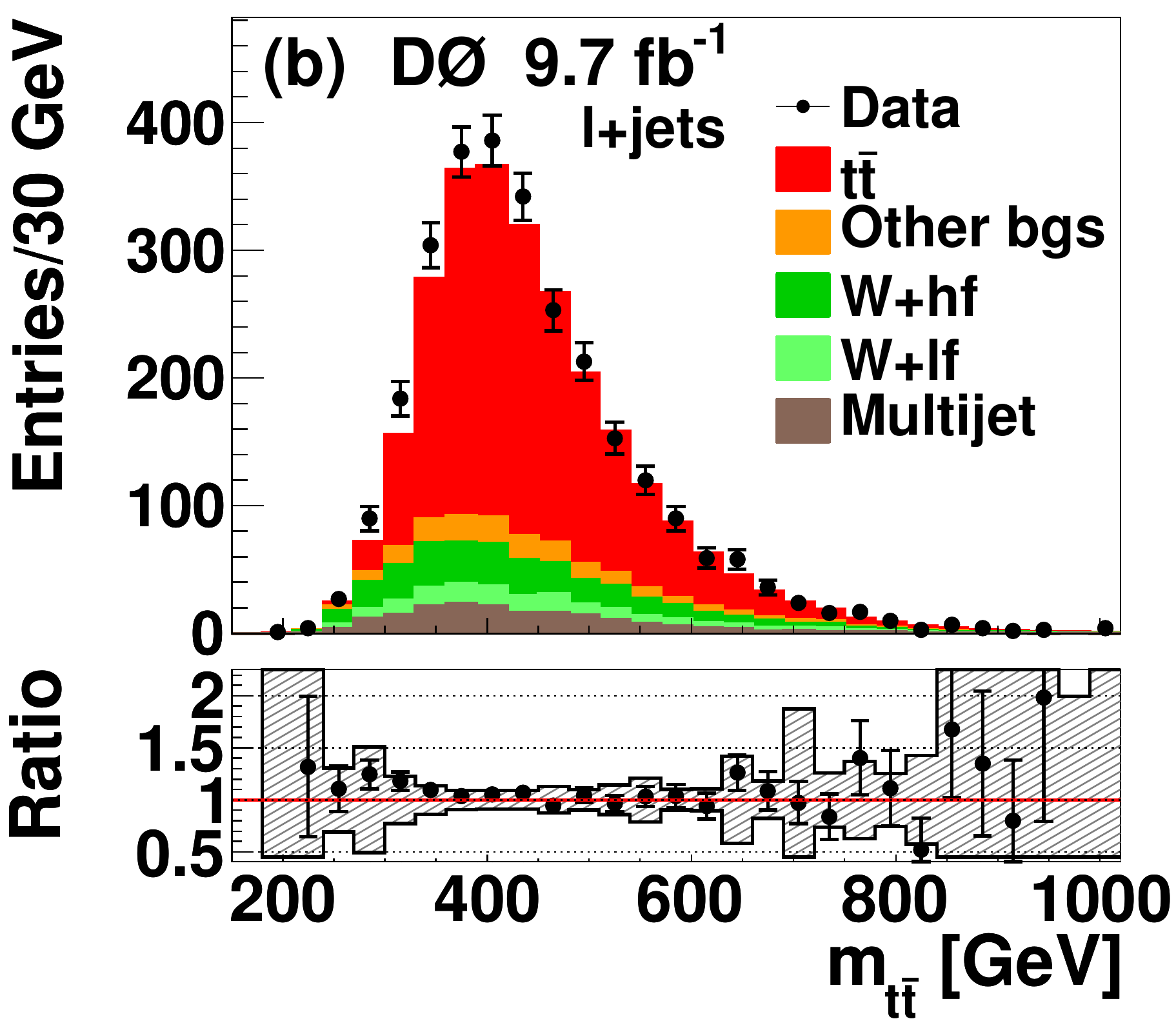}
\par\end{centering}
\caption{
\label{fig:sel}
(color online) 
(a) Invariant mass of the jet pair matched to one of the $W$ bosons.
(b) Invariant mass of the $\ttbar$ system. 
%The error bars on data points reflect the statistical uncertainty only. 
In the ratio of data to SM prediction, the total systematic uncertainty is shown as a shaded band. 
%For details about kinematic reconstruction and \ttbar signal normalization, see text.
}
\end{figure}

% calculation of P_bkg
The differential partonic cross section for \pbkg is calculated using the LO $W+4{\rm~jets}$ MEs implemented in \vecbos~\cite{bib:vecbos}. The initial-state partons are all assumed to have zero transverse momentum $\pt$.

Simulations are used to calibrate the ME technique. Signal \ttbar events, as well as the dominant background contribution from $W+{\rm jets}$ production, are generated with \alpgen~\cite{bib:alpgen} interfaced to \pythia. Therefore, it is the value of \mt as defined in the MC generator that is measured, and this value is expected to correspond within $\approx$ 1 GeV to \mt as defined in the pole mass scheme~\cite{bib:pdg}.
The detector response is fully simulated through {\sc geant3}~\cite{bib:geant}, followed by the same reconstruction algorithms as used on data.

% the calibration itself???
Seven samples of \ttbar events, five at $\mt^{\rm gen}=165, 170,$ $172.5, 175, 180~\GeV$ for $\kjes^{\rm gen}=1$, and two at $\kjes^{\rm gen}=0.95, 1.05$ for $\mt^{\rm gen}=172.5~\GeV$, are generated. Three samples of $W+{\rm jets}$ events, at $\kjes^{\rm gen}=0.95,1,$ and $1.05$, are produced. Together, the \ttbar, $W+{\rm jets}$ and MJ samples are used  to derive a linear calibration for the response of the ME technique to \mt and \kjes. For each generated $(\mt^{\rm gen},\kjes^{\rm gen})$ point, 1000 pseudo-experiments  (PE) are constructed, each containing the same number of events as observed in data.

\begin{table}
\begin{centering}
\begin{tabular}{lc}
\hline
\hline 
Source of uncertainty & Effect on \mt (GeV) \\
\hline
\emph{Signal and background modeling:} & \\
~~Higher order corrections & $+0.15$ \\
~~Initial/final state radiation & $\pm0.09$ \\
~~Hadronization and UE & $+0.26$ \\
~~Color reconnection & $+0.10$ \\
~~Multiple $p\bar p$ interactions & $-0.06$ \\
~~Heavy flavor scale factor & $\pm0.06$\\
~~$b$-jet modeling & $+0.09$ \\
~~PDF uncertainty & $\pm0.11$ \\
\emph{Detector modeling:} & \\
~~Residual jet energy scale & $\pm0.21$\\
~~Flavor-dependent response to jets & $\pm0.16$\\
~~$b$ tagging & $\pm0.10$\\
~~Trigger & $\pm0.01$\\
~~Lepton momentum scale & $\pm0.01$\\
~~Jet energy resolution & $\pm0.07$\\
~~Jet ID efficiency & $-0.01$\\
\emph{Method:} & \\
~~Modeling of multijet events & $+0.04$\\
~~Signal fraction & $\pm0.08$\\
~~MC calibration & $\pm0.07$\\
\hline 
{\em Total systematic uncertainty} & $\pm0.49$\\
{\em Total statistical uncertainty} & $\pm0.58$\\
{\em Total uncertainty} & $\pm0.76$\\
\hline
\hline
\end{tabular}
\par\end{centering}
\caption{
\label{tab:syst}
Summary of uncertainties on the measured top quark mass. The signs indicate the direction of the change in \mt when replacing the default by the alternative model.
%For the meaning of the signs, see text.
%Entries marked with a $^\star$ are consistent with 0, and so the uncertainty on the method calibration (0.01~GeV) is quoted instead.
}
\end{table}

% result: mt, kjes from data
Applying the ME technique to data, we measure after all calibrations
%\begin{equation}
$\mt=174.98\pm0.58~\GeV$ and $\kjes=1.025\pm0.005\,,$ 
%\end{equation}
where the total statistical uncertainty on \mt also includes the statistical contribution from \kjes. Splitting the total statistical uncertainty into two parts from \mt alone and \kjes, we obtain  $\mt=174.98\pm0.41\thinspace({\rm stat})\pm0.41\thinspace({\rm JES})~\GeV$. The two-dimensional likelihood distribution in $(\mt,\kjes)$ is shown in Fig.~\ref{fig:like2d}(a). Figure~\ref{fig:like2d}(b) compares the measured total statistical uncertainty on \mt with the distribution of this quantity from the PEs at $\mt^{\rm gen}=172.5~\GeV$ and $\kjes^{\rm gen}=1$.

Comparisons of SM predictions to data for $\mt=175~\GeV$ and $\kjes=1.025$ are shown in Fig.~\ref{fig:sel} for the invariant mass of the jet pair matched to one of the $W$ bosons and the invariant mass of the $\ttbar$ system.

% categorisation of systematic uncertainties
Systematic uncertainties are evaluated using PEs constructed from simulated signal and background events, for three categories:
modeling of signal and background events,
uncertainties in the simulation of the detector response,
and uncertainties associated with procedures used and assumptions made in the analysis. Contributions from these sources are listed in Table~\ref{tab:syst}.

The dominant category of systematic uncertainty is the modeling of signal events, with the largest contribution from {\em hadronisation and underlying event (UE)}, which is evaluated by comparing events simulated with \alpgen interfaced to either \pythia or \herwig. The JES calibration is derived using \pythia with a modified tune~A~\cite{bib:jes}, and is expected to be valid for this configuration only. Applying it to events that use \herwig for evolving parton showers can lead to a sizable effect on \mt.  However, this effect would not be present if the JES calibration were based on \herwig. To avoid such double-counting of uncertainty sources, we evaluate the uncertainty from hadronization and UE by considering as $\x$ the momenta of particle level jets matched in $(\eta,\phi)$ space to reconstructed jets. In this evaluation, we reweight our default \ttbar simulations in $\pt^{\ttbar}$ to match \alpgen interfaced to \herwig. 
Another important contribution to the systematic uncertainty is from {\em higher order corrections}, which is evaluated by comparing events simulated with {\sc mc@nlo}~\cite{bib:mcnlo} to \alpgen interfaced to \herwig~\cite{bib:herwig}. The uncertainty from the modeling of {\em initial and final state radiation} is constrained from Drell-Yan events~\cite{bib:isr}. As indicated by these studies, we change the amount of radiation via the renormalization scale parameter for the matching scale in \alpgen interfaced to \pythia~\cite{bib:isrmangano} up and down by a factor of 1.5. In addition, we reweight \ttbar simulations in $\pt$ of the \ttbar system ($\pt^{\ttbar})$ to match data, and combine the two effects in quadrature.

The category of systematic uncertainty from modeling of the detector response is dominated by the {\em residual jet energy scale} uncertainty from a potential dependence of the JES on $(\pt,\eta)$. Its impact on \mt is estimated by changing the jet momenta as a function of $(\pt,\eta)$ by the upper limits of JES uncertainty, the lower limits of JES uncertainty, and a linear fit within the limits of JES uncertainty. The maximum excursion in \mt is quoted as systematic uncertainty. Dedicated calibrations to account for the {\em flavour-dependent response} to jets originating from a gluon, a $b$ quark and $u,d,c,$ or $s$ quarks
are now an integral part of the JES correction~\cite{bib:jes}, and the uncertainty on \mt from these calibrations is evaluated by changing them within their respective uncertainties. This systematic uncertainty accounts for the difference in detector response to $b$- and light-quark jets.

In summary, we measure 
\begin{eqnarray*}
\mt &=& 174.98 \pm 0.58\thinspace({\rm stat+JES}) \pm 0.49\thinspace({\rm syst})~\GeV
\end{eqnarray*}
 with the ME technique in \ljets final states, which is consistent with the values given by the current Tevatron and world combinations of the top quark mass~\cite{bib:combitev,bib:combiworld} and achieves by itself a similar precision. With an uncertainty of $0.43\%$, it constitutes the most precise single measurement of \mt.

\section{Tevatron combination and outlook}

\begin{figure}
\begin{centering}
\includegraphics[width=0.99\columnwidth]{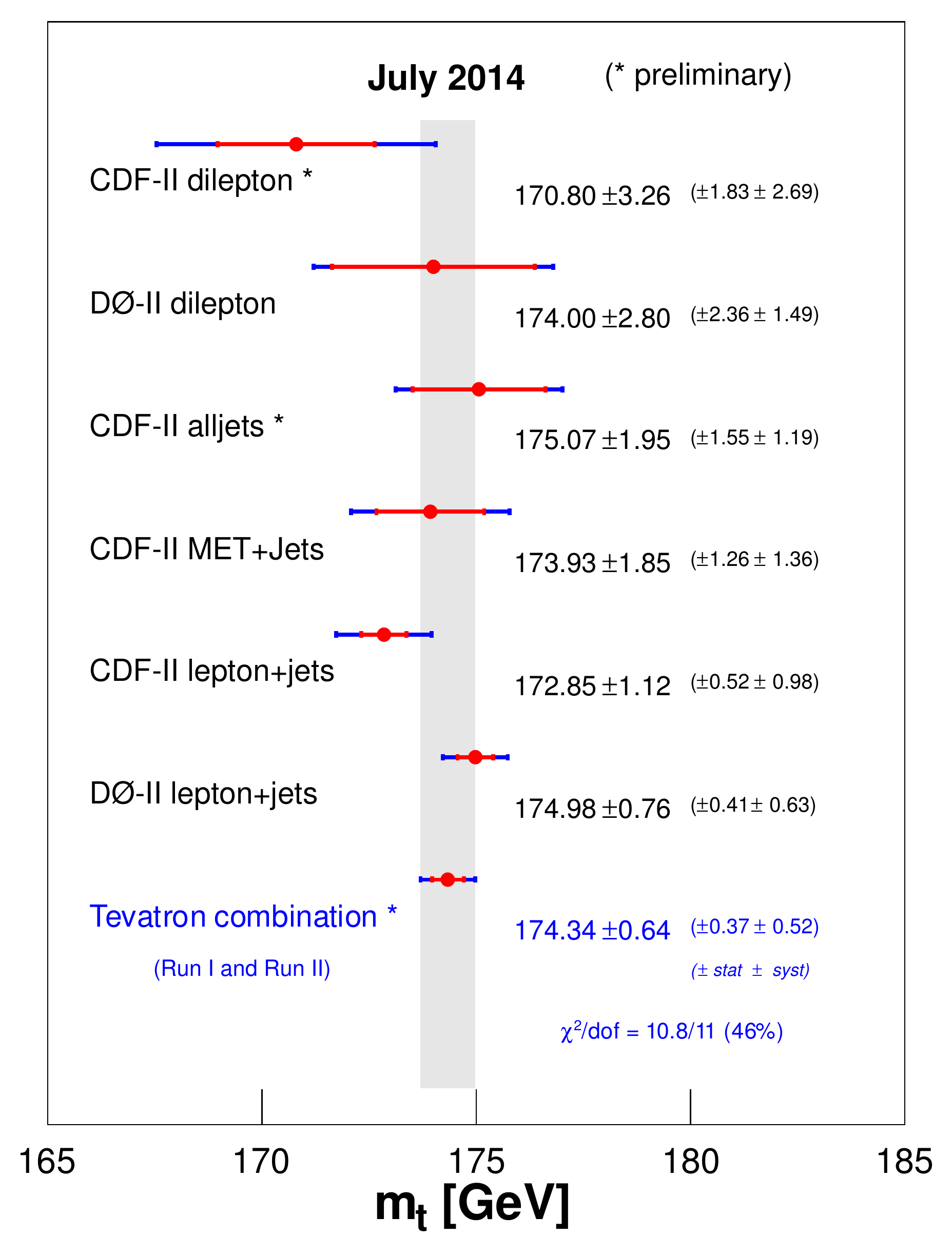}
\par\end{centering}
\caption{
\label{fig:combi}
Summary of the measurements performed in Run~II at the Tevatron which are used as inputs to the Tevatron combination~\cite{bib:combitev}. The inner uncertainty bars in red indicate the statistical uncertainty, while the outer uncertainty bars in blue represent the total uncertainty. The Tevatron average value of \mt obtained using input measurements from Run~I and Run~II is given at the bottom, and its uncertainty is shown as a gray band. The Figure is adapted from Ref.~\cite{bib:combitev}.
}
\end{figure}

Our results are included in the Tevatron combination from July 2014~\cite{bib:combitev}, which is performed taking into account 10 published and 2 preliminary results from the CDF and D0 collaborations using $p\bar p$ collision data from Run~I and Run~II of the Fermilab Tevatron Collider. Taking into account potential correlations between considered sources of systematic uncertainty as described in great detail in Ref.~\cite{bib:combitevprd}, the final result reads $\mt=174.34\pm0.64~\GeV$ corresponding to a precision of 0.37\%, with a relative contribution from our  measurement in \ljets final states of 67\%. An overview of input measurements performed using Run~II data is presented in Fig.~\ref{fig:combi}. The consistency of the input measurements is given by the value of the $\chi^2$ distribution for 11 degrees of freedom and corresponds to a $\chi^2$ probability of 46\%.

\section*{Acknowledgments}
I would like to thank my collaborators from the D0 experiment for their help in preparing this article. I also thank the staffs at Fermilab and collaborating institutions, as well as the D0 funding agencies.



%% References

\nocite{*}
\bibliographystyle{elsarticle-num}
\bibliography{martin}






% \end{thebibliography}

\end{document}